\documentclass{elsart}
\usepackage{epsfig}

\begin{document}
\begin{frontmatter}

\title{Experimental investigation of coherent Smith-Purcell radiation from a "flat" grating}

\author[Tomsk]{A. Aryshev\thanksref{someone}},
\author[Tomsk]{B. Kalinin},
\author[Tomsk]{G. Naumenko},
\author[Tomsk]{A. Potylitsyn},
\author[Moscow]{R. Bardai},
\author[Moscow]{B. Ishkhanov},
\author[Moscow]{V. Shvedunov}.
\address[Tomsk]{Institute for Nuclear Physics at Tomsk Polytechnic University,
634050, Pr. Lenina 2a, Tomsk, Russia.}
\address[Moscow]{Institute for Nuclear Physics of Moscow State University, Moscow, Russia.}
\thanks[someone]{Author to whom the correspondence should be addressed: A.S. Aryshev \\
Electronic address: alar@chair12.phtd.tpu.edu.ru}

\begin{abstract}
Using the pre-bunched electron beam of 5-MeV linear accelerator the
coherent Smith-Purcell radiation (CSPR) from a flat periodic target (made of conductive layers separated by dielectric gaps) in millimeter wavelength region has been investigated. The angular distribution of this radiation was measured with a narrow-band detector and
experimental data agree with our theoretical calculations for similar kind
of targets. Such properties of Smith-Purcell radiation as strong dependence
of radiation wavelength on the observation angle overlapping with coherent radiation effect may be used for a noninvasive bunch length measurement.
The possibility of using the room temperature detectors for single bunch measurements is demonstrated.
\end{abstract}


\end{frontmatter}

\section{Introduction}

The diffraction radiation (DR) is stipulated by means of dynamic perturbation of
electronic shells of atoms and free electrons by an electromagnetic field of
a charged particle moving near a conductive target. An effective transversal size of a
relativistic electron field for its component with wavelength $\lambda$ is about $\gamma \lambda$ where
$\gamma$ - is the Lorentz factor of an electron. This value can
reach macroscopic sizes in ultrarelativistic case, allowing the distance between an electron
beam and the target to be increased. A transition radiation, in contrast, is generated during passing of an initial particle through material of a target. In
case of the periodic target to be used, there maybe a constructive interference of
radiation from different elements of a target and the radiation becomes resonant
in this case. The special case of resonant diffraction radiation, when the
electron moves in parallel to a periodic target plane is referred as
Smith-Purcell radiation (SPR) \cite{1}. There are two types of targets for this
geometry of experiment: a volume grating obtained by a periodic deformation
of a continuous surface, and a flat periodic target obtained by alternation of metal strips and vacuum (dielectric) gaps. Authors of the work \cite{3} suggested to use CSPR for bunch length measurements. Changing the observation angle (in other words, the CSPR wavelength) one may detect a sharp increase in CSPR intensity when the wavelength becomes more than a bunch length. But the usage of a traditional lamellar grating for this aim has some disadvantages: for a set of observation angles there may exist deep minima in the CSPR angular distribution \cite{2}. It leads to significant troubles in interpretation of experimental data. Another method of the beam diagnostics suggested in \cite{4} is based on the rotation of the grating in the grating plane to change an effective grating period and needs the developed theory of SPR effect for a rotated grating that is absent now. 
As with of grating rotation the spectra of CSPR became wider and this technique should be investigated both theoretically and experimentally.
The use of a flat grating for the beam diagnostic technique allows one to avoid the angular distribution unevens and to increase the radiation yield \cite{5,6}. 

\section{Theoretical model}

The basis of theoretical model for calculation of radiation characteristics from a flat target is
the exact solution of Maxwell equations for diffraction radiation caused by
relativistic electron moving close to a thicknessless ideal conductive
half-plane \cite{7}.

As shown in \cite{5,8} the spectral - angular distribution of SPR from
a flat grating consisting of \textit{ N} strips with width\textit{ a} and period\textit{ d}, separated by vacuum gaps with width \textit{ (d-a)} can be written as:

\begin{eqnarray}
\frac{{d^{2}W_{SP}} }{{d\lambda d\Omega} } = 4\frac{{\alpha} }{{\pi
}}\frac{{\hbar c}}{{\lambda ^{2}}}\frac{{exp\left[ { - \frac{{4\pi
h}}{{\gamma \lambda} }} \right]}}{{1/\beta - cos {\theta}
}}sin\left[ {\frac{{\pi a}}{{\lambda} }\left( {cos\theta -
\frac{{1}}{{\beta} }} \right)} \right] \cdot F_{N}
\end{eqnarray}

Here $\alpha = {\textstyle{{1} \over {137}}}$ -is a fine
structure constant, $\hbar c = 0.2 \quad eV \cdot \mu m$- conversion constant,
$h$- distance between the particle trajectory and the target surface, $\theta $ -
outgoing photon angle, $\beta = \sqrt {1 - \gamma ^{ - 2}} $- velocity of
particle in units of light speed,$\gamma$ is Lorentz-factor, $\lambda$ - is SPR wavelength, $F_{N} = \frac{{sin^{2}\left( {N \cdot
\phi _{0} /2} \right)}}{{sin^{2}\left( {\phi _{0} /2} \right)}}$ -
coefficient, taking into account the interference of radiation from $N$
periodically spaced strips,

\[
\phi _{0} = \frac{{2\pi d}}{{\lambda} }\left( {cos\theta - \frac{{1}}{{\beta
}}} \right)
\]

Fig. 1 shows the calculated dependence of radiation wavelength on
the observation angle which follows from the equation for $\phi _{0} = 2\pi
k$ (Smith-Purcell dispersion law). It is evidently that, with increasing the
observation angle the spectral lines are shifted to large wavelength region.

With the radiation wavelength exceeding the longitudinal size of an electron
bunch, all the electrons from a bunch emit coherently. In this case radiation yield, normalized to one electron, increases proportionally to the number of
electrons in the bunch. The yield of coherent radiation for a bunch with
number of electrons $N_{e}$ in the aperture of the detector $\Delta \Omega $
registered by the detector with bandwidth $\Delta \lambda $ can be expressed as follows:

\begin{eqnarray}
\Delta W_{exp}\left( {h,\theta ,L_{bunch}}  \right) = N_{e} \cdot \int\limits_{\Delta
\lambda}  {\left| {f\left( {\frac{{\lambda} }{{L_{bunch}} }} \right)}
\right|^{2} \cdot \int\limits_{\Delta\Omega}  \frac{{d^{2}W_{SP}} }{{d\lambda d\Omega} }
d\Omega \cdot \mathop {\varepsilon \left( {\lambda}
\right)}\limits^{}}  \mathop {d\lambda} \limits^{}
\end{eqnarray}

Here $f\left( {\frac{{\lambda} }{{L_{bunch}} }} \right)$- is the
longitudinal form-factor of a bunch, $\varepsilon \left( {\lambda}  \right)$
- spectral efficiency of the detector, $L_{bunch}$ - length of the bunch. In Fig.~2 is given the relation between the squared form-factor modulus and radiation wavelengths for gauss longitudinal
distribution of electrons in a bunch. In figure we consider a bunchlength $L_{bunch}$ equal to $3\sigma $ in longitudinal distribution of electrons. It is clear from the figure that, for example, for $50\%$ level of a maximum intensity corresponds wavelength $\lambda =0.7 \cdot L_{bunch}$. That is, the measurement of angular distribution of CSPR (in other words, measurement of spectral dependence of a CSPR yield) allows a length of an electron bunch to be determined.

\section{Experimental setup}

The experiment was carried out in NPI MSU on a linac with the following
parameters of electron beam (see Table~1).

The experimental setup is shown in Fig.~3. The beam with energy 5MeV is
extracted through a 20 micrometer titanium foil and
moves at a distance $h=2$~mm from a periodic target.
\begin{flushright}
Table1
\end{flushright}
\begin{center}
\begin{tabular}{|l|c|}           \hline
Electron energy  & $E_{e}=5$ MeV\\  \hline
Bunch population & $10^{9}$\\       \hline
Bunch length     & $L_{bunch} \approx 1 \sim 2$ mm\\  \hline
Bunches in macropulse & 22\\        \hline
Macropulse duration & $\tau \approx 7$ ns\\ \hline
Distance between the target center and the detector & $R = 1$ m\\  \hline
\end{tabular}
\end{center}
The target, size $50 \times 50 \times 1.5$ mm consisting of 8 copper
strips with 3 mm width and $30$ micrometer thickness, located 3 mm apart on an insulating substrate, was placed in parallel to beam in the goniometer, by which it was possible to tune the parallelism between the beam trajectory and the target and change the impact parameter $h$.

The detector may be rotated on the table with radius $R=1$ m that allows one to change angle of photon detection. The center of rotation coincides with a target center. In the experiment we
registered CSPR together with transition radiation (TR) background from the exit window foil.

The detector was made on the basis of a room-temperature resonance funnel-shaped antenna
with the following characteristics: 
resonant wavelength $\lambda_{res}=7.2$~mm, a band width $\Delta \lambda \approx 2$ mm, sensitivity: $ \approx 0.18 $ V/W, response time $\tau _{D} =10$ ns. The polar angle acceptance $\Delta \theta \approx 12 mrad \approx 0.06 \gamma^{-1}$. Assuming that the dielectric constant of the target insulator substrate is the same as for vacuum, we have calculated the radiation power registered by the detector for our condition and $L_{bunch}=1.5$mm (see formula(2)) and obtained value $\Delta W_{exp}/\tau\approx 0.9 W$.

\section{Experimental results}

Figure 4 presents the calculated dependence of CSPR intensity on the angle
of detection in the real aperture of the detector with regard to the spectral
efficiency. The experimental relations between CSPR intensity and angle of
observation, the impact parameter and a current of an electron beam
are shown in Figures 5,6,7.

Comparing Figures 4 and 5 it is possible to notice the reasonable agreement between
the theoretical model and experiment. The monochromaticity of the main maximum is
defined by the number of the lattice periods and band pass of the detector
$\Delta \lambda/ \lambda > 1/N\sim12\% $.

The narrow peaks in Fig.~5 for angles $\theta = 56^{0}$ and $73^{0}$ can
be explained by resonances of a narrow-band detector. To get dependence
of CSPR intensity on the impact parameter, the detector was installed at
angle $\theta = 95^{0}$ that corresponded to the maximal value of CSPR.

In Fig.~6 the value $h = 0$ corresponds to a beam passing exactly
through a target and the dip at this value $h$ is explained by scattering the
electrons on target material.

The approximation of dependence of CSPR intensity on the impact parameter by
exponential function $f = F + a \cdot exp\left[ { - b \cdot h} \right]$,
where $b = {\textstyle{{4\pi}  \over {\gamma < \lambda >} }}$, gives the value
of mean detected wavelength of CSPR $<\lambda > = 7.9 \pm 1.3 mm$, that agrees well with resonant wavelength of the detector.

For confirming a coherent character of radiation we measured a dependence of
its intensity on pulse current of a macrobunch, $I_{acc} $, proportional to the
number of particles in the bunch. It canbe seen from Fig.~7 that for
$I_{acc}~<~I_{0}~\approx~12$~mA the dependence of CSPR yield is
$Y\sim I_{acc}^{c} $, where $c = 2.02 \pm 0.07$. With experimental
error accuracy, the dependence of CSPR intensity on the current of electron beam is
square-law, that confirms the coherent character of radiation. For values
$I_{acc} $, exceeding $12$mA detector starts to work in a nonlinear regime. It is necessary to notice that the  experiment was performed for
$I_{acc} < I_{0} $ condition.

\section{Conclusion}

In conclusion we may note that using $5$ MeV beam of a linac NPI MSU we
investigated characteristics of CSPR and showed, that the CSPR mechanism
provided the high monochromaticity of radiation even for a target with $8$ periods only and the aperture of the detector $~ \gamma^{-1}$, whereas
the continuous component in a spectrum connected with background processes
(TR from exit window, for instance) is small enough. Also it is necessary to pay attention to the fact, that for $\theta >>1/\gamma $ the CSPR yield practically does not depend on $\gamma $
(for $\gamma =10$ and for $\gamma =1000$ difference is a few percent). The
macropulse power of CSPR is proportional to $n \cdot N^{2}_{e}$, where $n$ - the number of bunches in macropulse, (for a linac NPI MSU estimation from (2) gives $P\sim n \cdot N^{2}_{e}\Delta W_{exp}/ \tau \sim 0.9W$). For such accelerators as
KEK-ATF and SLAC FFTB with number of electrons in a bunch $N_{e} \sim 10^{10}$,
this power level is reached for a single bunch. Thus, the received power level
demonstrates a possibility of measurement of a single bunch length using a flat grating and a set of detectors with responsivity: $\sim  0.1 V/W$, operating at room temperature and with different resonance wavelengths located at different
angles $\theta $. A resonance condition and requirement of coherency of a bunch
radiation define the wavelength range accessible to the bunch length measurement by this method: $\lambda  \approx d \approx L_{bunch}$. For
example, for $L_{bunch}=0.1$ mm,$ d \approx 0.1$ mm,$ \lambda  \approx
0.1$ mm one can use the detector for a far infrared range.

\newpage
\section*{Figure captions}

Figure 1: Calculated dependence of radiation wavelength on observation angle (d=6 mm, $\gamma = 10$, $k = 1,3...$- diffraction orders).

Figure 2: Dependence of square of a form-factor on radiation wavelength for gauss longitudinal distribution of electrons in a bunch.

Figure 3: The experiment scheme

Figure 4: The calculated relation between CSPR intensity and angle of detection $\theta$ in the real aperture of the detector with account of spectral
effectiveness.

Figure 5: Experimental dependence of CSPR intensity from observation angle.
Dotted curve is a continuous background ``substrate''.

Figure 6: CSPR intensity as a function of impact parameter

Figure 7: Experimental relation between CSPR intensity and an electron beam current.

\newpage
\begin{figure}[hb]
\centerline{
\leavevmode \epsfxsize=1.30\textwidth \epsfbox{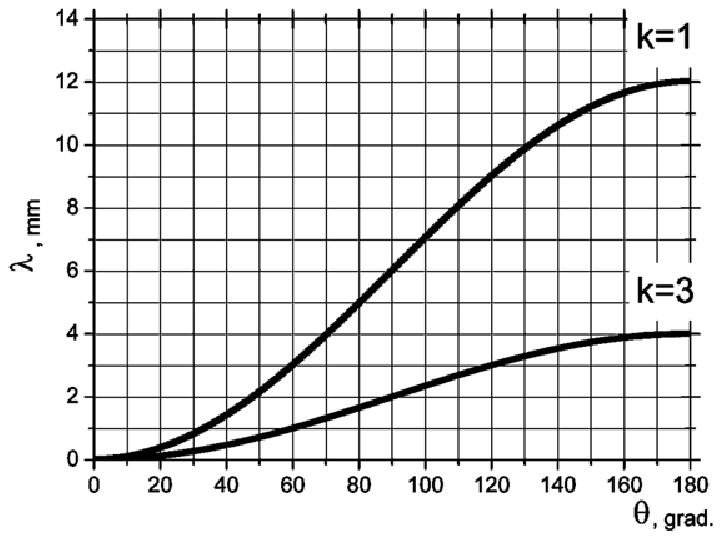}
}
\caption{}
\end{figure}

\newpage
\begin{figure}[hb]
\centerline{
\leavevmode \epsfxsize=1.30\textwidth \epsfbox{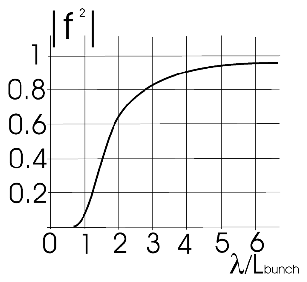}
}
\caption{}
\end{figure}

\newpage
\begin{figure}[hb]
\centerline{
\leavevmode \epsfxsize=1.30\textwidth \epsfbox{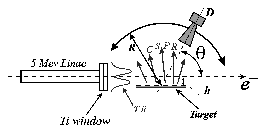}
}
\caption{}
\end{figure}

\newpage
\begin{figure}[hb]
\centerline{
\leavevmode \epsfxsize=1.30\textwidth \epsfbox{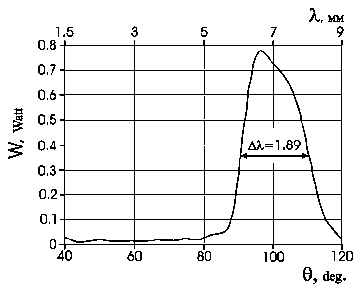}
}
\caption{}
\end{figure}

\newpage
\begin{figure}[hb]
\centerline{
\leavevmode \epsfxsize=1.30\textwidth \epsfbox{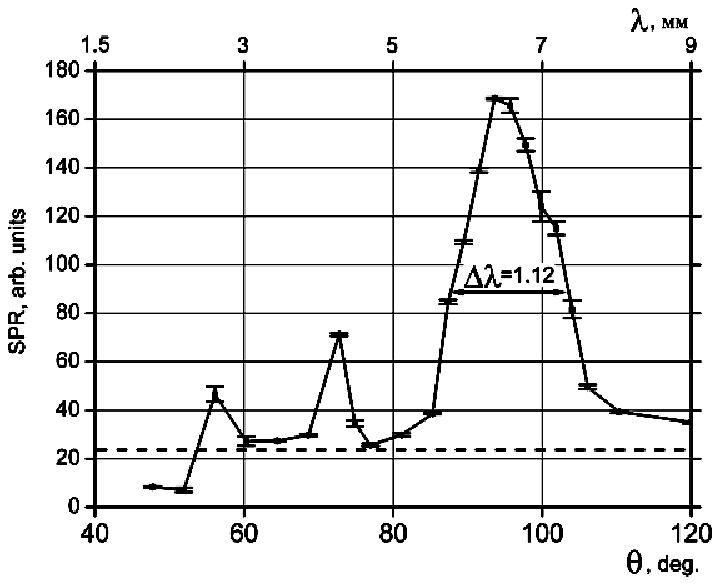}
}
\caption{}
\end{figure}

\newpage
\begin{figure}[hb]
\centerline{
\leavevmode \epsfxsize=1.30\textwidth \epsfbox{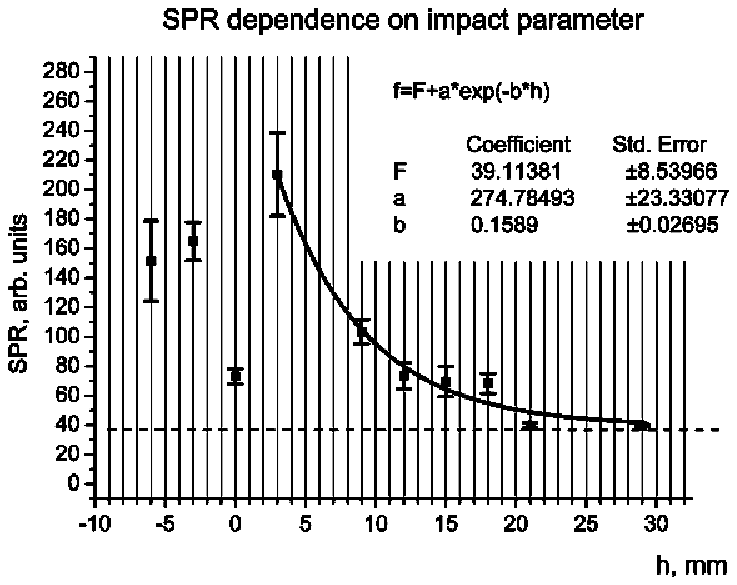}
}
\caption{}
\end{figure}

\newpage
\begin{figure}[hb]
\centerline{
\leavevmode \epsfxsize=1.30\textwidth \epsfbox{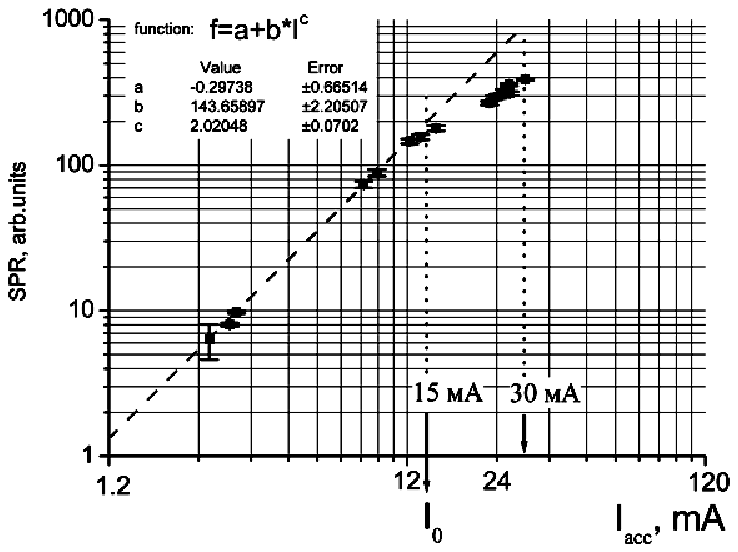}
}
\caption{}
\end{figure}

\end{document}